\documentstyle {article}

\def\t{\theta}
\def\g{\gamma}
\def\nn{\nonumber}
\def\ll{\lambda}

\def\N{\newline}
\def\o{\over}
\def\v{\varphi}
\def\p{\partial}
\def\l{\label}
\def\oms2{(1-\sqrt {2})}
\def\obs2{(1+\sqrt {2})}

\def\uv{(a_1-a_3)}
\def\uw{(a_1-a_4)}
\def\ux{(a_1-a_5)}
\def\uy{(a_1-a_6)}
\def\uz{(a_2-a_3)}
\def\vu{(a_2-a_4)}
\def\vv{(a_2-a_5)}
\def\vw{(a_2-a_6)}

\def\wz{(\oms2(a_1+a_3+a_4)-\obs2(a_2+a_5+a_6))}
\def\xu{(\oms2 (a_1+a_4+a_5)-\obs2 (a_2+a_3+a_6))}
\def\xv{(\oms2 (a_1+a_5+a_6)-\obs2 (a_2+a_3+a_4))}

\def\xz{(\oms2 (a_1+a_4+a_6)-\obs2 (a_2+a_3+a_5))}
\def\yu{(\oms2 (a_1+a_3+a_5)-\obs2 (a_2+a_4+a_6))}
\def\yv{(\oms2 (a_1+a_3+a_6)-\obs2 (a_2+a_4+a_5))}

\def\yx{(\obs2 (a_1+a_3+a_4)-\oms2 (a_2+a_5+a_6))}
\def\yy{(\obs2 (a_1+a_4+a_5)-\oms2 (a_2+a_3+a_6))}
\def\yz{(\obs2 (a_1+a_5+a_6)-\oms2 (a_2+a_3+a_4))}

\def\zx{(\obs2 (a_1+a_4+a_6)-\oms2 (a_2+a_3+a_5))}
\def\zy{(\obs2 (a_1+a_3+a_5)-\oms2 (a_2+a_4+a_6))}
\def\zz{(\obs2 (a_1+a_3+a_6)-\oms2 (a_2+a_4+a_5))}
\def\a{\alpha}

\def\be{\begin{equation}}
\def\bea{\begin{eqnarray}}
\def\ee{\end{equation}}
\def\eea{\end{eqnarray}}
\begin{document}
\begin{titlepage}
\begin{center}
\vspace*{1.5cm}
\hfill
\vbox{
    \halign{#\hfil         \cr
           hep-th/9710068 \cr
           IPM-97-241\cr
           October 1997\cr} 
      }  
\vskip 1cm
{\large \bf
Periods and Prepotential in $N=2$ Supersymmetric $E_6$
Yang-Mills Theory}
\vskip .6in
{\bf A.M. Ghezelbash}\footnote{e-mail:amasoud@physics.ipm.ac.ir}
\vskip .25in
{\em
Institute for Studies in Theoretical Physics and Mathematics, \\
P.O. Box 19395-5531, Tehran, Iran.\\
Department of Physics, Alzahra University, Vanak,
Tehran 19834, Iran.\\
}
\end{center}
\vskip 1.5cm
\begin{abstract}
We obtain the periods and one-instanton coefficient of
the $N=2$ supersymmetric Yang-Mills
theory with the exceptional gauge group $E_6$. These calculations are based on
the $E_6$ spectral curve and the obtained one-instanton coefficient is in
agreement with
the microscopic results.
\end{abstract}
\end{titlepage}\newpage
In the last few years, enormous advances have been made in understanding
of the low-energy behaviours of $N=2$ supersymmetric gauge theories.
Progress began with the paper of Seiberg and Witten \cite{1},
where the exact low-energy Wilsonian
effective action for the pure $N=2$ supersymmetric Yang-Mills theory with
the gauge group $SU(2)$ is derived. Since then, their work has been
generalized to other
pure gauge groups \cite{2} and to theories with the matter multiplet
\cite{3}.
In principle, the exact solution of such theories is given by an algebraic
curve. In the case of theories with classical Lie gauge groups, algebraic
curves are hyperelliptic \cite{4} which must be satisfied in the
consistency conditions of the relevant theories. The hyperelliptic curve of
the theories with the exceptional gauge groups are constructed in \cite{5,55}.
\newline
To understand the strong coupling region of the theory, one considers
the vevs of the Higgs fields $\vec a$ and their duals $\vec a_{D}$
which are related to the prepotential of the low-energy effective action,
$F(\vec a)$, by $\vec a_{D}={{\p F}\o {\p \vec a}}$.
These fields which are periods of the Riemann surface
$\pi =\pmatrix{{\vec a_{D}}\cr {\vec a}}$
, are represented by
the contour integrals of the Seiberg-Witten differential one-form,
$\ll$,
\be \l{AAD}
\vec a=\int _{\vec \alpha}\ll,\qquad\vec a_{D}=\int _{\vec \beta}\ll ,
\ee
where $\vec \alpha$ and $\vec \beta$ are homology cycles on the Riemann
surface.\N
To obtain the periods, $\vec a$ and $\vec a_D$, the
Picard-Fuchs (PF) operators, which annihilate $\vec a$ and $\vec a_{D}$
are usually used.
The PF equations have been derived in the case
of pure classical gauge groups and also classical gauge groups with
massless and massive supermultiplets \cite{6}.
In all these cases, the underlying algebraic curves of the theory are
hyperelliptic curves.
On the other hand, the algebraic curve for the supersymmetric
gauge theory is constructed from the spectral curve of the periodic Toda
lattice \cite{8}.
In the case of classical gauge groups, these curves are equivalent to the
hyperelliptic curves.
The PF equations for the supersymmetric $G_2$
gauge group has been constructed from the spectral curve of the
$(G_2^{(1)})^{\vee}$ Toda lattice theory \cite{9}.
Also it has been shown that the
calculation of n-instanton effects agree with the microscopic results
\cite{10}, while the calculation of one-instanton effect based on
the hyperelliptic curve shows different behaviour
with the microscopic results.\newline
Our motivation in this and subsequent works is to shed
light on the strong behaviour of $E_6$ theory and compare the
spectral and hyperelliptic curves of the
theory by finding the form of the periods and the
prepotential of the $E_6$ theory.
The PF equations of $E_6$ gauge theory based on the spectral curve
have been obtained in \cite{ITO,GH}.
In \cite{ITO}, PF equations are obtained from the topological two dimensional
Landau-Ginzburg theory, while in \cite{GH}, the PF equations are constructed
from the basic equations of the theory. However, a comparison of the PF
equations,
shows their equivalence.
In this paper,
we will give solutions of these
equations, and calculate the leading instanton corrections and find that the
one-instanton contribution of the prepotential coincides with the one obtained
by using the direct instanton calculation \cite{10}.\N
We use the spectral curve of $E_6$ given by \cite{11},
\be \l{YE6}
\zeta+{w\o \zeta}=-u_6+{{q_1+p_1\sqrt{p_2}}\o{x^3}},
\ee
which is obtained by degeneration of $K_3$ surface to an $E_6$ type
singularity. The polynomials $q_1$,$p_1$ and $p_2$ are given by,
\bea \l{Q1P1P2}
q_1&=&270 x^{15} + 342 u_1 x^{13} + 162 u_1^2 x^{11} - 252 u_2 x^{10} + (26
u_1^3 + 18 u_3) x^9
-162 u_1 u_2 x^8 + (6 u_1 u_3 - 27 u_4 ) x^7\nn\\&-&(30 u_1^2 u_2 - 36 u_5 ) x^6
+ (27 u_2^2 - 9 u_1 u_4 ) x^5 - (3 u_2 u_3 - 6 u_1 u_5 ) x^4 - 3 u_1 u_2^2 x^3
- 3 u_2 u_5 x -u_2^3,\nn\\
p_1&=&78 x^{10}+60 u_1 x^8 +14 u_1^2 x^6-33 u_2 x^5+2 u_3 x^4-5 u_1 u_2 x^3
- u_4 x^2 - u_5 x - u_2^2,\nn\\
p_2&=&12 x^{10}+12 u_1 x^8 +4 u_1^2 x^6-12 u_2 x^5+ u_3 x^4 - 4 u_1 u_2 x^3
-2 u_4 x^2 + 4 u_5 x + u_2^2,
\eea
where $u_1,\,u_2,\,u_3,\,u_4,\,u_5
$ and $u_6$ are the Casimirs of $E_6$.\N
We use the PF equations of \cite{ITO}, and by introducing the new variables,
\be \l{XS}
x_1={{u_4u_5}\o{u_2u_6}},\, x_2={{u_6}\o{u_1^2u_4}},\,x_3={{u_6}\o{u_1u_2^2}}
,\,x_4={{u_1u_3}\o{u_4}},\,x_5={{u_6}\o{u_3^2}},\,x_6={{w}\o{u_6^2}},
\ee
we get the following differential operators,
\bea \l{EQU}
E_1&=&32x_5\t _3 (\t _3-1)-6\t _{16}+3\t _{26}
-6\t _{36}+3\t _{56}-3\t _6,\nn\\
E_2&=&3x_1\t _{26}-\t _{45},\nn\\
E_3&=&4x_1x_4x_5\t _{23}-\t _{15},\nn\\
E_4&=&4x_4x_5\t _{34}-3\t _{16},\nn\\
E_5&=&x_1x_2x_4\t _{14}+8x_1x_4x_5\t _{3}(\t _3-1)+2\t _{35}+6x_1\t _{36},\nn\\
E_6&=&2x_1x_2x_4^2x_5\t _4 (\t _4-1)+12x_1x_4x_5\t _{36}+3\t _{56}
+9x_1\t _{6}(\t _6-1),\nn\\
E_7&=&x_3\t _2 (\t _2-1)-12x_2\t _{14}-18\t _{16}
-18\t _{26}+3x_2x_4\t _{4}(\t _4-1)-18\t _{46}-18\t _6,\nn\\
E_8&=&4x_1x_2\t _1 (\t _1-1)+8x_1x_4x_5\t _{13}-x_1x_2x_4\t _{14}
+\t _{15}+8x_1x_4x_5\t _{34}+2x_1^2x_2x_4^2x_5\t _{24}-\t _{25}
-\t _{45}\nn\\&+&12x_1x_4x_5\t _{3}+3x_1\t _{6},\nn\\
E_9&=&2x_1^2x_2x_4^2x_5\t _{24}+4x_1x_4x_5\t _{35}+\t _{5}(\t _5-1)+3x_1\t _{56}
,\nn\\
E_{10}&=&8\t _{13}-2x_4\t _{14}+x_4\t _{24}-2x_4\t _{34}+x_4\t _{45},\nn\\
E_{11}&=&4\t _1 (\t _1-1)+20\t _{12}+24\t _{13}+32\t _{14}+36\t _{15}+48\t _{16}
+25\t _2 (\t _2-1)+60\t _{23}+80\t _{24}+90\t _{25}+120\t _{26}\nn\\
&+&36\t _3 (\t _3-1)+96\t _{34}+108\t _{35}+144\t _{36}
+64\t _4 (\t _4-1)+144\t _{45}+192\t _{46}
+162\t_5 (\t _5-1)+216\t _{56}\nn\\&+&144(1-4x_6)\t _{6}(\t _6-1)+15\t _{2}+24\t _{3}
+48\t _{4}+63\t _{5}+120\t _{6}+1,\nn\\
E_{12}&=&x_3\t _{15}+6x_1x_2x_4^2x_5\t _4(\t _4-1)+18x_1^2x_2x_4^2x_5\t _{16},
\nn\\
E_{13}&=&36x_1x_2\t _{16}-x_3\t _{25}-12x_1x_2x_4^2x_5\t _4(\t _4-1)-9x_1x_2x_4
\t _{46},
\nn\\
E_{14}&=&2x_3\t _{35}+9x_1x_2x_4\t _{46}+27x_1^2x_2x_4\t _6(\t _6-1),
\nn\\
E_{15}&=&x_3\t _{5}(\t _5-1)-144x_1^2x_2x_4x_5\t _{36}+36x_1^2x_2x_4^2x_5
\t _{46}
+27\t _6(\t _6-1),
\nn\\
E_{16}&=&x_1x_2x_3x_4\t _{12}-9x_1x_2x_4(\t _{16}+\t _{26}+\t _{56}+\t _6)
-36x_1^2x_2x_4^2x_5\t _{36}+2x_3\t _{35}.
\eea
In the above equations, $\t _i$'s are the Euler derivatives
$\t _i=u_i{{\partial}\o{\partial u_i}}$ which are related to the Euler
derivatives
$\v _i=x_i{{\partial}\o{\partial x_i}}$ by the following relations,
\be \l{DER}
\t _1=-2\v _2-\v _3+\v _4,
\t _2=-\v _1-2\v _3,
\t _3=\v _4-2\v _5,
\t _4=\v _1-\v _2-\v _4,
\t _5=\v _1,
\t _6=-\v _1+\v _2+\v _3+\v _5-2\v _6,
\ee
and $\t _{ij}=\t _i\t _j$.\N
Now, we construct the solutions of the PF equations $E_i\pi=0\, (i=0,\cdots,16)$
in the semi-classical region where $w$ is small. We consider the following power
series solution around $(x_1,x_2,x_3,x_4,x_5,x_6)=(0,0,0,0,0,0)$,
\bea \l{SOL}
\rho _{{\alpha _1},{\alpha _2},{\alpha _3},{\alpha _4},{\alpha _5},{\alpha _6}}
(x_1,x_2,x_3,x_4,x_5,x_6)&=&\sum _{{m_1},{m_2},{m_3},{m_4},{m_5},{m_6}\geq0}
c_{{\alpha _1},{\alpha _2},{\alpha _3},{\alpha _4},{\alpha _5},{\alpha _6}}
(m_1,m_2,m_3,m_4,m_5,m_6)
\nn\\ & &x_1^{{m_1}+{\alpha _1}}x_2^{{m_2}+{\alpha _2}}
x_3^{{m_3}+{\alpha _3}}x_4^{{m_4}+{\alpha _4}}x_5^{{m_5}+{\alpha _5}}
x_6^{{m_6}+{\alpha _6}},
\eea
where $c_{{\alpha _1},\cdots ,{\alpha _6}}(0,\cdots, 0)=1$. By inserting eq.
(\ref{SOL}) in the eqs. (\ref{EQU}), we get the following indicial equations,
\bea \l{INDI}
&&4\ll _1 (\ll _1-1)+20\ll _1\ll _2+24\ll _1\ll _3+32\ll _1\ll _4+36\ll _1\ll _5
+48\ll _1\ll _6
+25\ll _2 (\ll _2-1)+60\ll _2\ll _3+80\ll _2\ll _4+90\ll _2\ll _5\nn\\
&+&120\ll _2\ll _6
+36\ll _3 (\ll _3-1)+96\ll _3\ll _4+108\ll _3\ll _5+144\ll _3\ll _6
+64\ll _4 (\ll _4-1)+144\ll _4\ll _5+192\ll _4\ll _6\nn\\
&+&162\ll _5 (\ll _5-1)+216\ll _5\ll _6+144\ll _{6}(\ll _6-1)+15\ll _{2}+24\ll
_{3}
+48\ll _{4}+63\ll _{5}+120\ll _{6}+1=0,\nn\\
&&\ll _6(2\ll _1-\ll _2+2\ll _3-\ll _5+1)=0,
\ll _6(\ll _1+\ll _2+\ll _4+1)=0,
\ll _5(\ll _1-\ll _2-\ll _4)=0,
\ll _5(\ll _5-1)=0,
\ll _4(\ll _4-1)=0,\nn\\
&&\ll _6(\ll _6-1)=0,
\ll _4\ll _5=0,\,   \ll _1\ll _5=0,\,  \ll _1\ll _6=0,\,\ll _3\ll _5=0,\,
\ll _5\ll _6=0,\,  \ll _1\ll _3=0,\,\ll _4\ll _6=0,
\eea
where
\be
\ll _1=-2\alpha _2-\a _3+\a _4,\ll _2=-\a _1-2\a _3,
\ll _3=\a _4-2\a _5, \ll _4=\a _1-\a _2-\a _4,
\ll _5=\a _1,\ll _6=-\a _1+\a _2+\a _3+\a _5-2\a _6.
\ee
The equations (\ref{INDI}) have the following solutions,
\bea \l{SOLUTION}
(\a _1,\cdots, \a _6)&=&(0,1/6,-1/6,-1/6,-1/12,-1/24),\nn\\
(\a _1,\cdots, \a _6)&=&(0,-1/6,1/2,1/6,-5/12,-1/24),\nn\\
(\a _1,\cdots, \a _6)&=&(0,\a _2,-3\a _2,-\a _2,2\a _2-1/12,-1/24),\nn\\
(\a _1,\cdots, \a _6)&=&(0,\a _2,-1/2\a _2-1/12,-\a _2,-1/2\a _2,-1/24),\nn\\
(\a _1,\cdots, \a _6)&=&(0,-2\a _3+5/6,\a _3,2\a _3-11/6,\a _3-11/12,-1/24),\nn
\\
(\a _1,\cdots, \a _6)&=&(0,\a _2,-3\a _2-1,-\a _2-1,2\a _2+11/12,-1/24).
\eea
The coefficients
$
c_{{\alpha _1},{\alpha _2},{\alpha _3},{\alpha _4},{\alpha _5},{\alpha _6}}
(m_1,m_2,m_3,m_4,m_5,m_6) $
obey the recursion relations,
\bea \l{REC}
c_{\vec{\a}}
(\vec{m})&=&
A_{\vec{\a}}
(\vec{m})
c_{\vec{\a}}
(m_1,m_2,m_3,m_4,m_5-1,m_6),\nn\\
c_{\vec{\a}}
(\vec{m}) &=&
B_{\vec{\a}}
(\vec{m})
c_{\vec{\a}}
(m_1-1,m_2,m_3,m_4,m_5,m_6),\nn\\
c_{\vec{\a}}
(\vec{m})&=&
C_{\vec{\a}}(\vec {m})
c_{\vec{\a}}(m_1-1,m_2,m_3,m_4-1,m_5-1,m_6),\nn\\
c_{\vec{\a}}
(\vec{m})&=&
D_{\vec{\a}}
(\vec{m})
c_{\vec{\a}}
(m_1,m_2,m_3,m_4-1,m_5-1,m_6),\nn\\
c_{\vec{\a}}
(\vec{m})&=&
E_{\vec{\a}}
(\vec{m})
c_{\vec{\a}}
(m_1-1,m_2-1,m_3,m_4-1,m_5,m_6)
+F_{\vec{\a}}
(\vec{m})
c_{\vec{\a}}
(m_1-1,m_2,m_3,m_4-1,m_5-1,m_6)\nn\\
&+&G_{\vec{\a}}
(\vec{m})
c_{\vec{\a}}
(m_1-1,m_2,m_3,m_4,m_5,m_6),\nn\\
c_{\vec{\a}}
(\vec{m})&=&
H_{\vec{\a}}
(\vec{m})
c_{\vec{\a}}
(m_1-1,m_2-1,m_3,m_4-2,m_5-1,m_6)
+I_{\vec{\a}}
(\vec{m})
c_{\vec{\a}}
(m_1-1,m_2,m_3,m_4-1,m_5-1,m_6)\nn\\
&+&J_{\vec{\a}}
(\vec{m})
c_{\vec{\a}}
(m_1-1,m_2,m_3,m_4,m_5,m_6) ,\nn\\
c_{\vec{\a}}
(\vec{m}) &=&
K_{\vec{\a}}
(\vec{m})
c_{\vec{\a}}
(m_1,m_2,m_3-1,m_4,m_5,m_6)
+L_{\vec{\a}}
(\vec{m})
c_{\vec{\a}}
(m_1,m_2-1,m_3,m_4,m_5,m_6) \nn\\
&+&M_{\vec{\a}}
(\vec{m})
c_{\vec{\a}}
(m_1,m_2-1,m_3,m_4-1,m_5,m_6),\nn\\
c_{\vec{\a}}
(\vec{m}) &=&
N_{\vec{\a}}
(\vec{m})
c_{\vec{\a}}
(m_1-1,m_2-1,m_3,m_4,m_5,m_6)
+(O_{\vec{\a}}+Q_{\vec{\a}}+S_{\vec{\a}})
(\vec{m})
c_{\vec{\a}}
(m_1-1,m_2,m_3,m_4-1,m_5-1,m_6) \nn\\
&+&P_{\vec{\a}}
(\vec{m})
c_{\vec{\a}}
(m_1-1,m_2-1,m_3,m_4-1,m_5,m_6)
+
R_{\vec{\a}}
(\vec{m})
c_{\vec{\a}}
(m_1-2,m_2-1,m_3,m_4-2,m_5-1,m_6)\nn\\
&+&T_{\vec{\a}}
(\vec{m})
c_{\vec{\a}}
(m_1-1,m_2,m_3,m_4,m_5,m_6), \nn\\
c_{\vec{\a}}
(\vec{m}) &=&
U_{\vec{\a}}
(\vec{m})
c_{\vec{\a}}
(m_1-2,m_2-1,m_3,m_4-2,m_5-1,m_6)
+V_{\vec{\a}}
(\vec{m})
c_{\vec{\a}}
(m_1-1,m_2,m_3,m_4-1,m_5-1,m_6) \nn\\
&+&W_{\vec{\a}}
(\vec{m})
c_{\vec{\a}}
(m_1-1,m_2,m_3,m_4,m_5,m_6),\nn\\
c_{\vec{\a}}
(\vec{m}) &=&
X_{\vec{\a}}
(\vec{m})
c_{\vec{\a}}
(m_1,m_2,m_3,m_4-1,m_5,m_6),\nn\\
c_{\vec{\a}}
(\vec{m}) &=&
Y_{\vec{\a}}
(\vec{m})
c_{\vec{\a}}
(m_1,m_2,m_3,m_4,m_5,m_6-1),
\eea
where the functions $A_{\vec{\a}}(\vec{m}),\cdots ,Y_{\vec{\a}}(\vec{m})$
are given in the appendix.
Therefore from these recursion relations, one can find the coefficients
$c_{\vec{\a}}(\vec{m})$, which some of them are as follows,
\bea \l{RECSOL}
c_{\vec{\a}}(m_1,0,0,0,0,0)=\prod _{i=1}^{m_1}B_{\vec{\a}}(i,0,0,0,0,0)&,&
c_{\vec{\a}}(0,m_2,0,0,0,0)=\prod _{i=1}^{m_2}L_{\vec{\a}}(0,i,0,0,0,0)\nn\\
c_{\vec{\a}}(0,0,m_3,0,0,0)=\prod _{i=1}^{m_3}K_{\vec{\a}}(0,0,i,0,0,0)&,&
c_{\vec{\a}}(0,0,0,m_4,0,0)=\prod _{i=1}^{m_4}X_{\vec{\a}}(0,0,0,i,0,0)\nn\\
c_{\vec{\a}}(0,0,0,0,m_5,0)=\prod _{i=1}^{m_5}A_{\vec{\a}}(0,0,0,0,i,0)&,&
c_{\vec{\a}}(0,0,0,0,0,m_6)=\prod _{i=1}^{m_6}Y_{\vec{\a}}(0,0,0,0,0,i)\nn
\eea
\bea \l{RECSOL2}
&&c_{\vec{\a}}(m_1,m_2,0,0,0,0)=N_{\vec{\a}}(m_1-1,m_2-1,0,0,0,0)
N_{\vec{\a}}(m_1-2,m_2-2,0,0,0,0)
\cdots N_{\vec{\a}}(\{\mid m_1-m_2\mid ,0\},0,0,0,0)\times \nn\\
&&c_{\vec{\a}}(\{\mid m_1-m_2 \mid ,0\},0,0,0,0)+\sum _{i=0}
^{{\rm min\{m_1,m_2\}}}\prod _{j=1}^{i} N_{\vec{\a}}(m_1-j,m_2-j,0,0,0,0)\times
\nn\\
&&T_{\vec{\a}}(m_1-i-1,m_2-i-1,0,0,0,0)c_{\vec{\a}}(m_1-i-1,0,0,0,0,0),\nn\\
&&c_{\vec{\a}}(m_1,0,m_3,0,0,0)=\prod _{i=1}^{m_1}Z_{\vec{\a}}(i,0,m_3+1,0,0,0)
c_{\vec{\a}}(0,0,m_3,,0,0,0),
\eea
where if $m_1 > m_2$, $\{\mid m_1-m_2 \mid ,0\}=m_1-m_2,0$ and else
$\{\mid m_1-m_2 \mid ,0\}=0,m_2-m_1$ and similar complicated structure
for the other coefficients. The function $Z_{\vec{\a}}$ is given in the appendix
.
To obtain the other solutions of eqs. (\ref{EQU}), we apply the well
known Frobenius
method, and find the logarithmic solutions of (\ref{EQU}). Hence the solutions
of the PF equations (\ref{EQU}) are given by,
\bea \l{SOLUTIONS}
\rho ^{(1)}_{\vec a}(\vec x)&=&\rho _{
(0,1/6,-1/6,-1/6,-1/12,-1/24)}(\vec x),\nn\\
\rho ^{(2)}_{\vec a}(\vec x)&=&\rho _{
(0,-1/6,1/2,1/6,-5/12,-1/24)}(\vec x),\nn\\
\rho ^{(3)}_{\vec a}(\vec x)&=&\rho _{
(0,\a _2,-3\a _2,-\a _2,2\a _2-1/12,-1/24)}(\vec x),\nn\\
\rho ^{(4)}_{\vec a}(\vec x)&=&\rho _{
(0,\a _2,-1/2\a _2-1/12,-\a _2,-1/2\a _2,-1/24)}(\vec x),\nn\\
\rho ^{(5)}_{\vec a}(\vec x)&=&\rho _{
(0,-2\a _3+5/6,\a _3,2\a _3-11/6,\a _3-11/12,-1/24)}(\vec x),\nn\\
\rho ^{(6)}_{\vec a}(\vec x)&=&\rho _{
(0,\a _2,-3\a _2-1,-\a _2-1,2\a _2+11/12,-1/24)}
(\vec x),\nn\\
{\rho ^{(1)}_{D}}_{\vec a}(\vec x)&=&({{\p}\o{\p \a _2}}-{{\p}\o{\p \a _3}}
-{{\p}\o{\p \a _4}}+{1\o 2}{{\p}\o{\p \a _5}}+{1\o 4}
{{\p}\o{\p \a _6}})\rho _{\vec \a}
(\vec x)\mid _{\vec \a=(0,1/6,-1/6,-1/6,-1/12,-1/24)},\nn\\
{\rho ^{(2)}_{D}}_{\vec a}(\vec x)&=&({{\p}\o{\p \a _2}}-3{{\p}\o{\p \a _3}}
-{{\p}\o{\p \a _4}}+{5\o 2}{{\p}\o{\p \a _5}}+{1\o 4}{{\p}\o{\p \a _6}})
\rho _{\vec \a}
(\vec x)\mid _{\vec \a=(0,-1/6,1/2,1/6,-5/12,-1/24)},\nn\\
{\rho ^{(3)}_{D}}_{\vec a}(\vec x)&=&({{\p}\o{\p \a _2}}-3{{\p}\o{\p \a _3}}
-{{\p}\o{\p \a _4}}+{5\o 2}{{\p}\o{\p \a _5}}+{1\o 4}
{{\p}\o{\p \a _6}})\rho _{\vec \a}
(\vec x)\mid _{\vec \a=(0,\a _2,-3\a _2,-\a _2,2\a _2-1/12,-1/24)},\nn\\
{\rho ^{(4)}_{D}}_{\vec a}(\vec x)&=&({{\p}\o{\p \a _2}}
-{{\p}\o{\p \a _4}}-{1\o 2}{{\p}\o{\p \a _5}}+1/4{{\p}\o{\p \a _6}})
\rho _{\vec \a}
(\vec x)\mid _{\vec \a=(0,\a _2,-1/2\a _2-1/12,-\a _2,-1/2\a _2,-1/24)},\nn\\
{\rho ^{(5)}_{D}}_{\vec a}(\vec x)&=&({{\p}\o{\p \a _3}}
+{1\o 4}{{\p}\o{\p \a _6}})\rho _{\vec \a}
(\vec x)\mid _{\vec \a=(0,-2\a _3+5/6,\a _3,2\a _3-11/6,\a _3-11/12,-1/24)},\nn
\\
{\rho ^{(6)}_{D}}_{\vec a}(\vec x)&=&({{\p}\o{\p \a _2}}
+{1\o 4}{{\p}\o{\p \a _6}})\rho _{\vec \a}
(\vec x)\mid _{\vec \a=(0,\a _2,-3\a _2-1,-\a _2-1,2\a _2+11/12,-1/24)}.
\eea
To find the classical solutions, we use the following relations,
\bea \l{RELCLAS}
u_1&=&-a_1^2-a_2^2-a_3^2-a_4^2-a_5^2-a_6^2+a_5a_4+a_1a_2+a_6a_3+a_3a_2
+a_4a_3 ,\nn\\
u_2&=&-a_1a_2a_3^2a_4+a_2a_5a_3^2a_4+a_1^2a_3^2a_4+a_1^2a_2a_4^2+a_1^2a_4^2a_5
-a_1^2a_4a_5^2-a_1a_2^2a_4^2+
a_1a_2^2a_6^2-a_1a_2^2a_5^2-a_1^2a_2a_6^2\nn\\&+&a_1^2a_2a_5^2+a_1^2a_6^2a_3
-a_1^2a_3a_4^2-a_1^2a_3^2a_6+a_2^2a_4^2
a_5-a_2^2a_4a_5^2+a_2^2a_3a_5^2-a_1^2a_2a_4a_3+a_1^2a_2a_6a_3-a_1^2a_2a_5a_4
\nn\\&+&a_1a_2a_5^2a_4-a_1a_2a_5a_4^2-a_1
a_2^2a_6a_3+a_1a_2^2a_4a_3+a_1a_2^2a_4a_5-a_1a_2a_6^2a_3+a_1a_2a_3a_4^2
+a_1a_2a_3^2a_6+a_2a_5^2a_3a_4\nn\\&-&a_2a_5
a_3a_4^2-a_2^2a_3a_5a_4-a_5a_6^2a_4^2+a_5^2a_4a_6^2-a_5^2a_3a_6^2
+a_5^2a_3^2a_6+a_5a_6^2a_4a_3-a_5^2a_6a_4a_3
+a_5a_6a_4^2a_3-a_5a_3^2a_6a_4\nn\\&-&a_2a_5^2a_3^2,
\eea
and complicated expressions for $u_3,u_4,u_5,u_6$ which we do not
express here. These relations are obtained through the following relation
between the classical part of hyperelliptic curve of ref. \cite{55} and the
functions $p_1,p_2,q_1$ given in (\ref{Q1P1P2}),
\be \l{REL}
-108x^3W(x)=-108x^3\prod _{i=1}^{27}(x-{\tilde a}_i)=(x^3u_6-q_1)^2-p_1^2p_2
\ee
where ${\tilde a}_i$ are given in ref. \cite{55}.
From the (\ref{RELCLAS}), one may construct the classical solutions,
\bea \l{CLASIC}
a_1&=&{3\o {\sqrt 2}}\rho ^{(4)}_{(0,-1/6,0,1/6,1/12,-1/24)}(\vec x),\nn\\
a_2&=&{{\sqrt 5}\o 2}\rho ^{(3)}_{(0,1/30,-1/10,-1/30,-1/60,-1/24)}(\vec x)+
{3\o {\sqrt 5}}\rho ^{(2)}_{\vec \a}(\vec x),\nn\\
a_3&=&{1\o {\sqrt 3}}\rho ^{(1)}_{\vec \a}(\vec x)+
{3\o {\sqrt 2}}\rho ^{(4)}_{(0,-1/6,0,1/6,1/12,-1/24)}(\vec x),\nn\\
a_4&=&\rho ^{(2)}_{\vec \a}(\vec x)+
{{\sqrt{5}}\o 3}\rho ^{(6)}_{(0,-1/4,-1/4,-3/4,5/12,-1/24)}(\vec x),\nn\\
a_5&=&
{1\o \sqrt{2}}\rho ^{(1)}_{\vec \a}(\vec x)+{{\sqrt 2}\o 3}
\rho ^{(5)}_{(0,-17/30,7/10,-13/30,-13/60,-1/24)}(\vec x)+\nn\\
&&{{\sqrt{2}}\o 3}\rho ^{(6)}_{(0,-1/4,-1/4,-3/4,5/12,-1/24)}(\vec x),\nn\\
a_6&=&{{\sqrt{3}}\o 3}\rho ^{(6)}_{(0,-1/4,-1/4,-3/4,5/12,-1/24)}(\vec x)
+{4\o {\sqrt 2}}\rho ^{(4)}_{(0,-1/6,0,1/6,1/12,-1/24)}(\vec x),\nn\\
{a_D}_1&=&{{2i}\o {\pi}}(\sqrt 3{\rho ^{(3)}_D}_{(0,1/30,
-1/10,-1/30,-1/60,-1/24)}(\vec x)
-{1\o 3}\sqrt 2{\rho ^{(5)}_D}_{(0,-17/30,7/10,-13/30,-13/60,-1/24)}(\vec x))+
\sum _{i=1}^6 \epsilon _{1i}\rho ^{(i)}_{\vec \a}(\vec x),\nn\\
{a_D}_2&=&{{i}\o{\pi}}(
{1\o 2}\sqrt 5{\rho ^{(2)}_D}_{\vec \a }(\vec x)+
{1\o 3}\sqrt 5{\rho ^{(4)}_D}_{(0,-1/6,0,1/6,1/12,-1/24)}(\vec x))+
\sum _{i=1}^6 \epsilon _{2i}\rho ^{(i)}_{\vec \a}(\vec x),\nn\\
{a_D}_3&=&{{2i}\o{\pi}}(
\sqrt 5{\rho ^{(2)}_D}_{\vec \a }(\vec x)+
{1\o 5}\sqrt 2{\rho ^{(4)}_D}_{(0,-1/6,0,1/6,1/12,-1/24)}(\vec x))+
\sum _{i=1}^6 \epsilon _{3i}\rho ^{(i)}_{\vec \a}(\vec x),\nn\\
{a_D}_4&=& {{4i}\o{\pi}}\sqrt 2{\rho ^{(1)}_D}_{\vec \a }(\vec x)+
\sum _{i=1}^6 \epsilon _{4i}\rho ^{(i)}_{\vec \a}(\vec x),\nn\\
{a_D}_5&=&{{i}\o {3\pi}}({1\o 2}{\rho ^{(5)}_D}_{
(0,-17/30,7/10,-13/30,-13/60,-1/24)}(\vec x)+
\sqrt 2{\rho ^{(1)}_D}_{\vec \a }(\vec x))+
\sum _{i=1}^6 \epsilon _{5i}\rho ^{(i)}_{\vec \a}(\vec x),\nn\\
{a_D}_6&=&{{2i}\o{\pi}}(
{{\sqrt 3}\o 3}{\rho ^{(4)}_D}_{(0,-1/6,0,1/6,1/12,-1/24)}(\vec x)+
{{\sqrt 2}\o 4}{\rho ^{(6}_D}_{(0,-1/4,-1/4,-3/4,5/12,-1/24)}(\vec x))+
\sum _{i=1}^6 \epsilon _{6i}\rho ^{(i)}_{\vec \a}(\vec x),\nn\\
\eea
where $\epsilon _{ij}$ are constants and are determined by the evaluating
of the period integrals. However, for the computation of the instanton
correction to the prepotential, explicit form of the ${a_D}_i$ is not
necessary.
From these solutions, we get the following identities,
\bea \l{I}
\sum _{i=1}^6 ({\p _{u_1}}{{a_D}_i} a_i-{{a_D}_i}{\p _{u_1}}a_i)&=&
{{12i}\o{\pi}}\nn\\
\sum _{i=1}^6 ({\p _{u_j}}{{a_D}_i} a_i-{{a_D}_i}{\p _{u_j}}a_i)&=&
0,\, j\in\{2,\cdots,6\}.
\eea
Although the proof of the above equations is difficult, we have explicitly
checked
(\ref {I}) up to order $w$. By integrating the identities (\ref {I}) over
$u_1,\cdots,u_6$, we get the scaling equation,
\be \l{SCAL}
\sum _{i=1}^6a_i {{\p {\cal F}}\o{\p a_i}}-2{\cal F}={{12i}\o{\pi}}u_1.
\ee
From the above equation and the following form of the prepotential $F({\vec a})$
in the semi-classical region,
\be \l{PRE}
F({\vec a})={{i}\o{4\pi}}\sum _{\a \in {\Delta_+}(E_6)}(\a ,a)^2\log
{{(\a ,a)^2} \o {w^{1/12}}}+1/2\tau _0\sum _{i=1}^6 (a_i)^2+\sum _{n=1}^\infty
{\cal F}_n({\vec a})w^n,
\ee
one can find the n-instanton coefficients ${\cal F}_n({\vec a})$. We get
the following form for the one-instanton effect which because of its complexity,
only one term of it is given explicitly here,
\bea \l{INS1}
{\cal F}_1({\vec a})&=&{{150}\o{i\pi}}\{ \uv \uw \ux \uy \uz \vu \vv \vw \nn\\
&&\wz \xu \nn\\
&&\xv \xz \nn\\
&&\yu \yv \nn\\
&&\yx \yy \nn\\
&&\zx \zy \nn\\
&&\zz \}/
\{(a_1-a_2)^2(a_1-a_3)^2(a_1-a_4)^2(a_1-a_5)^2(a_1-a_6)^2 \nn\\
&&(a_2-a_3)^2(a_2-a_4)^2(a_2-a_5)^2(a_2-a_6)^2{\wz }^2\nn\\
&&{\xu }^2{\xv }^2\nn\\
&&{\xz }^2{\yu }^2\nn\\
&&{\yv }^2{\yx }^2\nn\\
&&{\yy }^2{\yz }^2\nn\\
&&{\zx }^2{\zy }^2\nn\\
&&{\zz }^2\}+\cdots
\eea
Redefininhg $w$ as,
\be
w_{PV}={{1}\o {2^13 75}}w,
\ee
we find that one-instanton coefficient (\ref{INS1}) is in agreement with the
microscopic calculation of \cite{10}.\N
For comparison the strong behaviour of theory based on the hyperelliptic curve,
one must construct the corresponding PF equations and solve them, which will
appear in a forthcoming paper.
\N
\vspace*{5mm}
{\large ACKNOWLEDGEMENTS}
\vspace*{5mm}

The author would like to thank M. Alishahiha
for helpful discussions.

\vspace*{5mm}
{\large APPENDIX}
\vspace*{5mm}

Here, we list the functions $A_{\vec{\a}}(\vec{m}),\cdots
,Y_{\vec{\a}}(\vec{m})$
and $Z_{\vec{\a}}(\vec{m})$ which appear in the equations (\ref{REC}) and
(\ref{RECSOL2}),
\bea
A_{\vec{\a}}(\vec{m})={{32(\g _3+2)(\g _3+1)}\o{3\g _6(2\g _1-\g_2+2\g _3
-\g _5+1)}}&,&
B_{\vec{\a}}(\vec{m})={{3(\g _2+1)(\g _6+1)}\o{\g _4\g _5}},\nn\\
C_{\vec{\a}}(\vec{m})={{4(\g _2+1)(\g _3+1)}\o{\g _1\g _5}}&,&
D_{\vec{\a}}(\vec{m})={{4(\g _3+1)(\g _4+1)}\o{3\g _1\g _6}},\nn\\
E_{\vec{\a}}(\vec{m})={{-(\g _1+1)(\g _4+1)}\o{2\g _3\g _5}}&,&
F_{\vec{\a}}(\vec{m})={{-4(\g _3+1)}\o{\g _5}},\nn\\
G_{\vec{\a}}(\vec{m})={{-3(\g _6+1)}\o{\g _5}}&,&
H_{\vec{\a}}(\vec{m})={{-2(\g _4+1)(\g _4+2)}\o{3\g _5\g _6}},\nn\\
I_{\vec{\a}}(\vec{m})=F_{\vec{\a}}(\vec{m})&,&
J_{\vec{\a}}(\vec{m})=G_{\vec{\a}}(\vec{m}),\nn\\
K_{\vec{\a}}(\vec{m})={{(\g _2+1)(\g _2+2)}\o{18\g _6(\g _1+\g _2)}}&,&
L_{\vec{\a}}(\vec{m})={{-2(\g _1+2)(\g _4+1)}\o{3\g _6(\g _1+\g _2)}},\nn\\
M_{\vec{\a}}(\vec{m})={{(\g _4+1)(\g _4+2)}\o{6\g _6(\g _1+\g _2)}}&,&
N_{\vec{\a}}(\vec{m})={{4(\g _1+1)(\g _1+2)}\o{\g _5(-\g _1+\g _2+\g _4}},\nn\\
O_{\vec{\a}}(\vec{m})={{8(\g _1-1)(\g _3+1)}\o{\g _5(-\g _1+\g _2+\g _4}}&,&
P_{\vec{\a}}(\vec{m})={{-(\g _1+1)(\g _4+1)}\o{\g _5(-\g _1+\g _2+\g _4}},\nn\\
Q_{\vec{\a}}(\vec{m})={{8\g _4(\g _3+1)}\o{\g _5(-\g _1+\g _2+\g _4}}&,&
R_{\vec{\a}}(\vec{m})={{2(\g _2+2)(\g _4+1)}\o{\g _5(-\g _1+\g _2+\g _4}},\nn\\
S_{\vec{\a}}(\vec{m})={{12(\g _3+1)}\o{\g _5(-\g _1+\g _2+\g _4}}&,&
T_{\vec{\a}}(\vec{m})={{3(\g _6+1)}\o{\g _5(-\g _1+\g _2+\g _4}},\nn\\
U_{\vec{\a}}(\vec{m})={{2(\g _2+2)(\g _4+1)}\o{\g _5(-\g _5+1)}}&,&
V_{\vec{\a}}(\vec{m})=F_{\vec{\a}}(\vec{m}),\nn\\
W_{\vec{\a}}(\vec{m})=G_{\vec{\a}}(\vec{m})&,&
X_{\vec{\a}}(\vec{m})={{(\g _4+1)\{2(\g _1+\g _3)-\g _2-\g _5-4\}}
\o{8\g _1\g _3}},\nn
\eea
\bea
Y_{\vec{\a}}(\vec{m})&=&\{576(\g _6+1)(\g _6+2)\}/\{
4\g _1 (\g _1-1)+20\g _1\g _2+24\g _1\g _3+32\g _1\g _4+36\g _1\g _5+48\g _1\g _6
+25\g _2 (\g _2-1)\nn\\&+&60\g _2 \g _3+80\g _2\g _4+90\g _2\g _5+120\g _2\g _6
+36\g _3 (\g _3-1)+96\g _3\g _4+108\g _3\g _5+144\g _3\g _6
+64\g _4 (\g _4-1)\nn\\&+&144\g _4\g _5+192\g _4\g _6
+162\g _5 (\g _5-1)+216\g _5\g _6+144\g _6(\g _6-1)+15\g _{2}+24\g _{3}
+48\g _{4}+63\g _{5}+120\g _{6}+1\},\nn\\
Z_{\vec{\a}}(\vec{m})&=&{{-(\g _1+1)(\g _4+1)}\o{\g _5(-\g _1+\g _2+\g _4)}}.
\eea
In the above equations, $\g _i$ has been obtained from $\t _i$ given in
(\ref{DER}),
with the replacement
$\v _i \rightarrow m_i+\a _i$.

\end{document}